# Magnetic Structure of the Quasi-One-Dimensional $La_3OsO_7$ as Determined by Neutron Powder Diffraction

Ryan Morrow,[1] Michael A. Susner,[2,3] Michael D. Sumption,[2] and Patrick M. Woodward[1]

[1]Department of Chemistry and Biochemistry, The Ohio State University, Columbus, Ohio 43210-1185, USA

[2]Department of Materials Science and Engineering, The Ohio State University, Columbus, Ohio 43210-1185, USA

[3]Materials Science and Technology Division, Oak Ridge National Laboratory, Oak Ridge, TN 37831, USA

## Abstract

Insulating $5d^3$ $La_3OsO_7$ and hole doped $La_{2.8}Ca_{0.2}OsO_7$ materials featuring well separated pseudo-one-dimensional zig-zag chains of corner-sharing $OsO_6$ octahedra have been synthesized and their magnetic and electrical transport properties characterized. Long range magnetic order between the antiferromagnetic chains is determined with a propagation vector k = 1/2, 1/2, 0 and $T_N$ = 45 and 53 K for the parent and doped materials. An $Os^{5+}$ moment of 1.7(1) $\mu_B$ for $La_3OsO_7$ and 1.2(2) $\mu_B$ for $La_{2.8}Ca_{0.2}OsO_7$ is refined. The long range magnetic structure is compared to the few currently known for isostructural $Ln_3MO_7$ compounds.

## Introduction

Materials having quasi-one-dimensional (1D) structures have been widely studied for their interesting electronic and magnetic properties, such as charge density waves [1, 2], spin-Peierls transitions [3, 4], and novel magnetic excitations [5, 6]. The magnetism of several pseudo-1D structure types containing 3d transition metal ions have been studied [7, 8]. However, fewer materials have been investigated which exhibit one-dimensional features and contain 4d or 5d transition metal ions. Given the decreased electron correlations and increased spin orbit coupling of the 5d oxides the magnetism may be distinct from similar 3d oxides and merits a closer inspection.

One such system is a particular polymorph of the $Ln_3MO_7$ family which can be described as an ordered variant of the defect fluorite structure wherein Ln represents a trivalent rare earth cation and M is a pentavalent cation. M can be Sb [9], Nb [9, 10], Mo [11], Ru [12-14], Ta [9], Re [15, 16], Os [17-20], and Ir [21, 22], representing a considerable degree of chemical diversity. A central structural feature of these compounds is the formation of chains of corner sharing $MO_6$ octahedra which are isolated from one another and significantly kinked with bond angles between octahedra of approximately 145 degrees. Possibly due to the zig-zag configuration of these octahedral chains, the electrons are localized on the M cation and therefore enable the one-dimensional magnetic behavior in these compounds. The wide variety of M cations possible in this system provides for the opportunity to study one-dimensional magnetism with electron configurations ranging from $d^1$ to $d^4$.

While many previous reports have included studies of magnetic or structural transitions in these compounds via susceptibility or X-ray diffraction measurements, respectively, little work has been done with neutron diffraction to elucidate the magnetic structures of this interesting materials family. In the spin 3/2 compound $La_3RuO_7$, an ideal candidate for the study of the quasi-one-dimensional magnetism due to the non-magnetic rare earth cation choice, Khalifah *et al.* [23] reported evidence of magnetic neutron scattering at low temperatures. Though these data indicated the presence of long range magnetic order, they were not able to determine the magnetic structure. In the 5d analogue $La_3OsO_7$, Lam *et al.* [17] reported magnetic susceptibility data that differed considerably from that of $La_3RuO_7$, suggesting the two isoelectronic compounds may have different magnetic structures. However, neutron diffraction studies were not carried out. Here we report a neutron diffraction study of the crystal and magnetic structure of $La_3OsO_7$, as well as investigate the effects of hole doping through calcium substitution on the lanthanum site.

## Experimental

Polycrystalline samples of a maximum size of 1.6 g were prepared by combining $La_2O_3$ (99.99%, GFS, dried overnight at 1000°C prior to use) and Os metal (99.98%, Alfa Aesar) in an alumina vessel which was sealed in an evacuated silica tube (volume of approximately 40 mL with 3 mm thick walls) along with a secondary crucible containing $PbO_2$ which served as a sacrificial $O_2$ source at elevated temperatures. The reaction is thus described as:

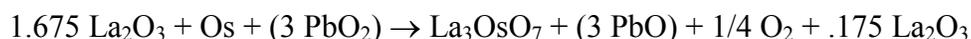
$1.675\ La_2O_3 + Os + (3\ PbO_2) \rightarrow La_3OsO_7 + (3\ PbO) + 1/4\ O_2 + .175\ La_2O_3$

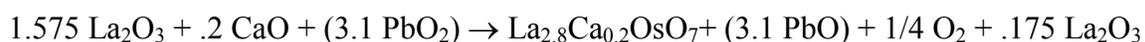
$1.575\ La_2O_3 + .2\ CaO + (3.1\ PbO_2) \rightarrow La_{2.8}Ca_{0.2}OsO_7 + (3.1\ PbO) + 1/4\ O_2 + .175\ La_2O_3$

In accordance with previous work on $La_3RuO_7$ [12], a calculated excess of $La_2O_3$ was used in order to prevent the formation of La deficient La-Os-O secondary phases. Although this was not reported to be necessary in the previous report of $La_3OsO_7$ [17], here it was found to be useful, particularly in the doped case. A calculated excess of $PbO_2$ was used in order to fully oxidize the target phase. Preparations were heated to 1000 °C for a period of 48 hours in a box furnace located in a fume hood. Care must be taken when heating osmium or osmium-containing compounds due to the potential formation of toxic volatile $OsO_4$. Additional care must be taken when evolving gas *in situ*, as tube rupture may occur.

Phase purity was verified via X-ray powder diffraction with a Bruker D8 Advance laboratory diffractomer equipped with a Ge (111) monochromator. Neutron powder diffraction data was collected at Oak Ridge National Laboratory's Spallation Neutron Source (SNS) on the POWGEN [24] beamline. Data were collected using the orange cryostat sample environment with 6 mm vanadium sample cans using chopper settings and wavelength ranges corresponding to bank 2 (0.2760–4.6064 Å) and bank 5 (2.2076–15.3548 Å). Data were collected for 4 hours, split between the banks at 75 and 165 minutes each respectively, at each measurement temperature – 5, 30, 70 and 300 K for $La_3OsO_7$ and 5, 25, 45, 70 and 300 K for $La_{2.8}Ca_{0.2}OsO_7$. Rietveld analysis was conducted using the software package GSAS EXPGUI [25, 26] on both banks simultaneously. Representational analysis was performed with SARAh [27], and magnetic form factor coefficients for Os were adapted from Kobayashi *et al.* [28].

Conductivity measurements were collected using a Quantum Design PPMS with a 4 point contact geometry on sintered bar shaped pellets. Data were collected with zero applied field between the temperatures 300 K and 150 K, below which the samples became too resistive to measure accurately. Magnetic measurements were conducted on Quantum Design SQUID MPMS and VSM PPMS devices using loose polycrystalline powders and sintered pellets, respectively, in the temperature ranges 2 to 400 K under a variety of applied fields with FC and ZFC conditions upon warming.

## Results

The neutron powder diffraction data were able to be fit readily with the *Cmcm* spacegroup, using the starting model that had previously been determined [17]. This crystal structure, which has been previously described [12], is shown in Figure 2 and consists of chains of $OsO_6$ octahedra connected at transverse corners aligned parallel to the *c* axis. The chains are significantly buckled, with bond angles of approximately 147° kinking in alternating orientations along the *b* axis. The distance between Os cations in neighboring chains is approximately 6.75 Å. Next nearest neighboring chains are separated by distances equivalent to the *a* and *b* lattice parameters, approximately 7.5 and 11.2 Å, respectively. Detailed refined structural parameters are available in Tables 1 and 2 for $La_3OsO_7$ and $La_{2.8}Ca_{0.2}OsO_7$ respectively. Refined anisotropic displacement parameters are available as Supplemental Material.

There is also evidence of the successful doping of Ca into the structure from the neutron powder diffraction data analysis. While the size of the $Ca^{2+}$ is quite similar to that of $La^{3+}$, it is slightly smaller (1.12 versus 1.16 Å) [29], and there is a small decrease in the unit cell volume in the doped compound. Additionally, no added impurity reflections are present in the Ca doped material. The refined weight percentages of the $La_2O_3$ impurity, which was added as a reagent with a predetermined excess, were 7.5(2) and 8.1(2)% for the parent and doped compound, respectively. These values are in close agreement with the values expected given the initial excesses used (7.3 and 7.5%). The occupancies were set to reflect the expected Ca content randomly distributed over both La sites. Attempts to refine the occupancies of these sites did not result in substantial changes.

The electrical conductivity of the sintered polycrystalline pellets, shown in Figure 3, displays clear insulating behavior with an increase in resistivity of approximately 3 orders of magnitude in the temperature range 150 to 300 K. There was a decrease of resistivity of approximately one half of an order of magnitude with doping; however, the insulating behavior remains intact. The data is shown to be linear when plotted on a $T^{-1/2}$ temperature scale, in accordance with a variable range hopping model [30] with a dimensionality of one ($T^{-1/(1+d)}$ where d is the dimensionality) with an $R^2$ of 0.9999. Although the data range being fit is not exhaustive, there is a noticeable preference to this model with respect to a simple activated model ($T^{-1}$; $R^2$ = 0.9985) and 3D variable range hopping model ($T^{-1/4}$; $R^2$ = 0.9991). The 1D variable range hopping mechanism can be easily rationalized given the one-dimensional nature of the corner sharing $OsO_6$ chains which are then disjointed and randomly oriented within the polycrystalline sample.

The temperature dependence of the magnetization of both compounds is shown in Figure 4. For $La_3OsO_7$ a broad feature near 100 K can be attributed to intrachain one-dimensional magnetic correlations. Although this feature is not as clear as has been previously shown [17], it is visible as a deviation from Curie-Weiss behavior. The very sharp cusp at 45 K is due to three dimensional magnetic ordering; an additional broad feature centered at 12 K is noted. All three features are consistent with the observations of Lam *et al.* [17]. The Ca doped material's magnetization temperature dependence has a number of features, including a similar broad feature near 100 K, with deviation from Curie-Weiss behavior at a higher temperature, and apparent phase transitions at 53, 33, and 19 K, the latter two having appreciable dependency with magnetic field. Both data sets could be fit to a Curie-Weiss law in the temperature range 200 – 400 K, further confirming the concept of localized magnetic moments. Effective moments of 3.14 and 2.83 $\mu_B$ were calculated for the parent and doped material, respectively, as well as respective Weiss constants of −204 and −200 K. The reduced effective moment supports the conclusion of oxidizing Os in the material by hole doping. Both of these effective moments are reduced from the theoretical spin only value of 3.87 $\mu_B$, however Lam *et al.* [17] performed a similar analysis with data collected up to 600 K, showing that the slope of the inverse magnetic susceptibility changed at high temperatures resulting in an effective moment within error of the theoretical spin-only result.

In $La_3OsO_7$, below the temperature of the sharp magnetic transition which has been previously assigned to a likely three-dimensional magnetic ordering, several additional magnetic reflections were detected in the neutron powder diffraction data (Figure 5). The two strongest magnetic reflections could be indexed as the (1/2, 1/2, 1) and (3/2, 1/2, 1) reflections, indicating a propagation vector of **k** = 1/2, 1/2, 0. Representational analysis using SARAh yielded two potential irreducible representations which dictated the orientation of the Os moments and coupling to neighboring Os moments within a given chain. The output of the SARAh program is given as Supplemental Material. Γ(1) contained basis vectors which would allow neighboring spins within a chain to have antiferromagnetically coupled components within the *ab* plane and ferromagnetically coupled *c* components. Γ(3) contained basis vectors which would allow neighboring spins within a chain to have ferromagnetically coupled components within the *ab* plane and antiferromagnetically coupled *c* components.

Attempts at fitting both of these possibilities resulted in the determination that Γ(1) was the correct choice, although the inclusion of a ferromagnetic *c* component was not supported by refinement attempts or the previously described magnetization measurements and was therefore fixed at zero. The refinement of this model had some sensitivity to the orientation of the moments within the *ab* plane with the best fit corresponding to a structure where the moments are tilted by 20-30 degrees from the *a* axis, approximately in the direction of the nearest neighbor chain along the face diagonal (approximately 33.8 degrees away from the *a* axis). A representation of the magnetic structure is shown in Figure 6. Each Os moment is coupled antiferromagnetically to the Os in the next nearest neighboring chains, along the *a* and *b* axes. Each Os moment has two antiferromagnetic pairs and two ferromagnetic pairs with the Os in the four nearest neighbor chains.

At the lowest temperature measured, the magnetic moment refined on $Os^{5+}$ was 1.7(1) $\mu_B$. This number is significantly reduced from the expectation of a magnetic ion with 3 unpaired electrons. However, the result is in excellent agreement with several recent neutron diffraction studies in $Os^{5+}$ bearing double perovskites [31, 32] where the moment is found to be reduced primarily due to the effects of a high degree of covalency in the Os−O bonds. There did not appear to be any substantial change in the magnetic structure between 5 and 30 K that would give some clues about the origin of the lower temperature (12 K) feature in the magnetic susceptibility data. There is a small increase in the intensity of the (1/2, 1/2, 1) peak relative to (3/2, 1/2, 1) which may correspond to a slight reorientation of the moment in the *ab*-plane as the refined moment angle suggests, but the refinement was not sensitive enough to definitively link this subtle change with the feature in the susceptibility data.

In considering the magnetic structure refinement of the hole doped $La_{2.8}Ca_{0.2}OsO_7$, there were no obvious features in the low temperature neutron diffraction data to suggest that the nature of the magnetic structure had changed from the parent compound save for an overall reduction in intensity of the magnetic scattering presumably resulting from the reduction of the number of unpaired electrons from 3 to 2.8. The moment refined on Os was 1.2(2) $\mu_B$ at 5 K which is a greater reduction in moment than might be expected from the absence of 0.2 electrons per Os. Unfortunately, there were no clues from the neutron diffraction data collected at temperatures between the cusps seen in the susceptibility data to offer conclusive evidence as to their nature, except that the onset of long range ordering of the moments can be associated with the 53 K transition.

It is instructive to consider the magnetic structure of $La_3OsO_7$ in the context of prior magnetic neutron scattering work on this family available in the literature. In the very similar material $La_3RuO_7$, only two magnetic reflections were observed [23] with d spacings of 9.80 Å and 12.20 Å. These positions do not correspond to the scattering which would result from the application of the magnetic structure of $La_3OsO_7$ to that system. In fact, one must at least quadruple the unit cell in all three dimensions to find possible indices of the peaks observed in $La_3RuO_7$. These deviations indicate that the magnetic structure of $La_3RuO_7$ is most certainly different than what is found for $La_3OsO_7$ studied here, as had been predicted [17]. In the study of the $d^1$ $La_3MoO_7$ [11], a small increase in the intensity of several nuclear peaks was observed and attributed to a 3D ordered magnetic structure, presumably with a propagation vector of $\mathbf{k} = 0, 0, 0$. This intensity was fit with a model wherein nearest-neighbor antiferromagnetic chains are antiferromagnetically ordered, which does not require an expansion of the unit cell to describe. Using this language, the magnetic structure of $La_3OsO_7$ could then be described as an antiferromagnetic ordering of the next-nearest-neighbor chains. It is remarkable that three distinct 3D magnetic structures appear to have been observed in these three compounds with electronic configurations of $4d^1$, $4d^3$, and $5d^3$ and nonmagnetic La cations. Furthermore, it has been observed in the $4d^3$ cases of $Nd_3RuO_7$ and $Pr_3RuO_7$ that ordering schemes corresponding to nearest neighbor AFM Ru chain ordering [33] and an as-yet-unsolved magnetic structure with a propagation vector $\mathbf{k} = 1/2, 1/2, 1/2$ are possible [34], indicating that a magnetic rare earth cation can affect the magnetic structure of the chains in a substantial way. It would be interesting to further understand how additional electronic configurations in the $La_3MO_7$ family might order, what the true magnetic structure of $La_3RuO_7$ is, and how the addition of a magnetic Ln cation affects the long range magnetic ordering of the chains.

In conclusion, the 3D ordered magnetic structure of the quasi-one-dimensional $La_3OsO_7$ compound has been determined as next nearest neighbor antiferromagnetic ordering of antiferromagnetic chains from neutron powder diffraction and compared to the current state of knowledge of magnetic structures for this class of materials. Additionally, the effects of hole doping through Ca substitution for La have been explored without clear evidence of remarkable impact on the magnetic or electrical properties. Several low temperature features have been observed in magnetic susceptibility of the doped material which appear to be strongly affected by the strength of an applied magnetic field and merit further study.

## Supplemental Material

A refined bank 5 diffraction pattern, refined anisotropic displacement parameters, a magnetic scattering figure for $La_{2.8}Ca_{0.2}OsO_7$, and the SARAh output are available as Supplemental Material.

## Author Information

**Corresponding Author**

Morrow.176@osu.edu

**Notes**

The authors declare no competing financial interest.

## Acknowledgements

Support for this research was provided by the Center for Emergent Materials an NSF Materials Research Science and Engineering Center (DMR-0820414). Additional support was provided by the U.S. Department of Energy, Office of High Energy Physics under Grant Number DE-FG02-95ER40900. A portion of this research was carried out at Oak Ridge National Laboratory's Spallation Neutron Source, which is sponsored by the U.S. Department of Energy, Office of Basic Energy Sciences. The authors thankfully acknowledge Pamela Whitfield for experimental assistance with POWGEN data collection.

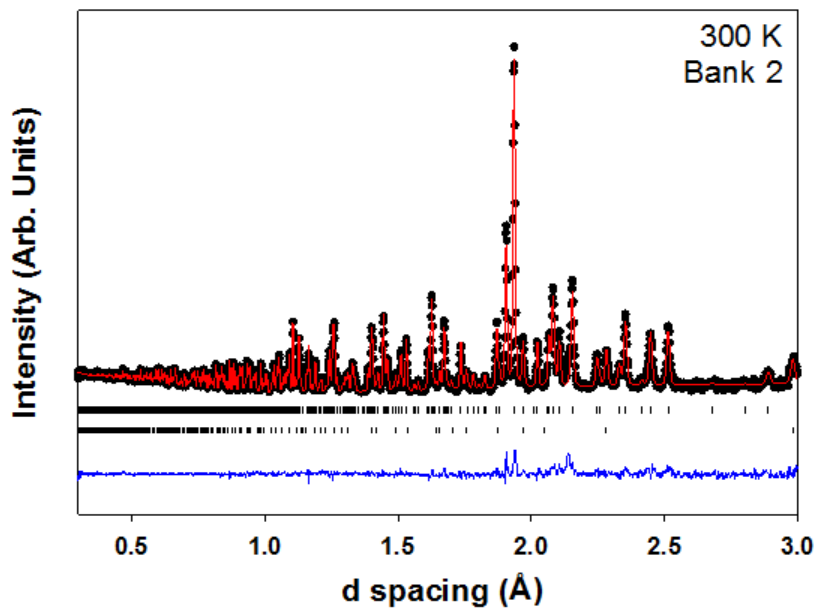

FIG. 1 (color online). Refined neutron powder diffraction data for $La_3OsO_7$ at 300 K. Black symbols, red curves, and blue curves represent the observed data, calculated pattern, and difference respectively. The upper and lower sets of hash marks signify the main phase and excess $La_2O_3$ impurity, respectively.

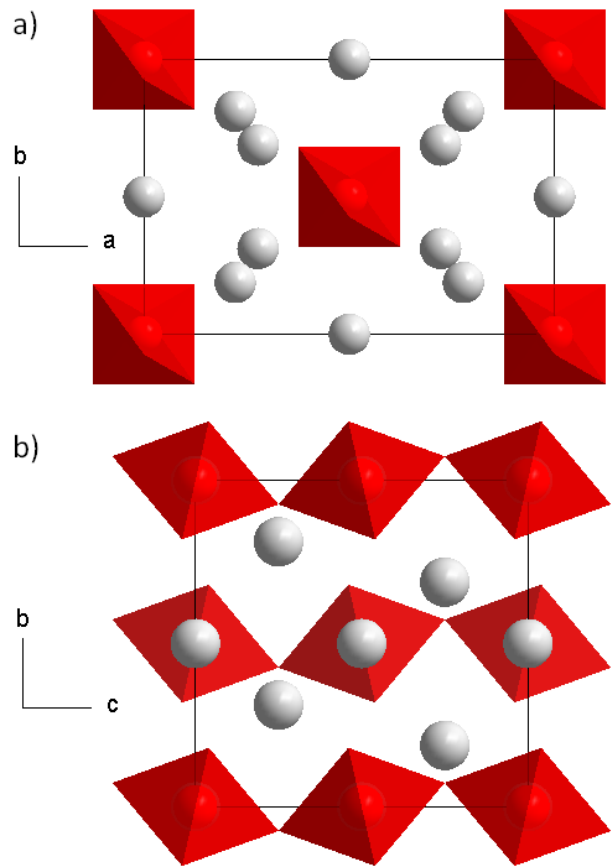

FIG. 2 (color online). Crystal structure of $La_3OsO_7$ shown with views a) down the $c$ axis and b) down the $a$ axis revealing the buckling of the chains of $OsO_6$ octahedra. $OsO_6$ are shown as red octahedra with Os spheres occupying the center and O located at the vertices, La ions are shown as grey spheres, and additional O anions are omitted for clarity.

|  | La$_3$OsO$_7$ | | | |
| --- | --- | --- | --- | --- |
|  | 5 K | 30 K | 70 K | 300 K |
| Space Group | *Cmcm* | *Cmcm* | *Cmcm* | *Cmcm* |
| a (Å) | 11.1966(2) | 11.1960(3) | 11.1963(3) | 11.2009(3) |
| b (Å) | 7.4915(1) | 7.4920(2) | 7.4960(2) | 7.5224(2) |
| c (Å) | 7.6198(1) | 7.6192(2) | 7.6185(2) | 7.6210(2) |
| V (Å)$^3$ | 639.15(1) | 639.10(4) | 639.41(5) | 642.13(4) |
| $R_{wp}$ | 3.60% | 3.64% | 3.63% | 3.29% |
| $\chi^2$ | 2.864 | 2.928 | 2.918 | 2.449 |
| Os-O1 (×2) (Å) | 1.98812(2) | 1.98728(4) | 1.98823(4) | 1.9907(4) |
| Os-O2 (×4) (Å) | 1.96848(2) | 1.96854(3) | 1.96778(3) | 1.963(1) |
| ∠Os-O1-Os | 146.739(1) | 146.872(1) | 146.653(1) | 146.3 (1) |
| La2 x | 0.2774(1) | 0.2773(1) | 0.2774(1) | 0.2776(1) |
| La2 y | 0.3119(2) | 0.3121(2) | 0.3121(2) | 0.3118(2) |
| O1 y | 0.4240(3) | 0.4244(3) | 0.4239(3) | 0.4233(4) |
| O2 x | 0.1239(1) | 0.1238(1) | 0.1240(1) | 0.1234(1) |
| O2 y | 0.1820(2) | 0.1819(2) | 0.1814(2) | 0.1808(2) |
| O2 z | 0.0403(2) | 0.0405(2) | 0.0405(2) | 0.0404(2) |
| O3 x | 0.3697(1) | 0.3697(1) | 0.3697(1) | 0.3695(1) |
| O3 y | 0.0317(2) | 0.0315(2) | 0.0315(2) | 0.0314(2) |
| La1 U | 0.0030(2) | 0.0038(3) | 0.0044(3) | 0.0103(3) |
| La2 U | 0.0015(2) | 0.0017(2) | 0.0021(2) | 0.0053(2) |
| Os U | 0.0015(2) | 0.0014(2) | 0.0020(2) | 0.0042(2) |
| O1 $U_{eq}$ | 0.0041(5) | 0.0035(5) | 0.0037(5) | 0.0072(6) |
| O2 $U_{eq}$ | 0.0062(3) | 0.0066(3) | 0.0073(3) | 0.0147(4) |
| O3 $U_{eq}$ | 0.0033(3) | 0.0034(4) | 0.0035(4) | 0.0063(4) |
| Os moment ($\mu_B$) | 1.7(1) | 1.7(1) | - | - |
| Moment angle | 21(7) | 27(8) | - | - |

TABLE 1 – Refinement parameters from neutron powder diffraction from La$_3$OsO$_7$. $U_{eq}$ is defined as a third of the trace of the tensor for anisotropically refined U's, which are available as Supplemental Material. Moment angle is defined as the angle away from the *a* axis within the *ab* plane. Special positions used are La1 (0, 1/2, 0), La2 (x, y, 1/4), Os (0, 0, 0), O1 (1/2, y, 1/4), and O3 (x, y, 1/4).

$La_{2.8}Ca_{0.2}OsO_7$

| | 5 K | 25 K | 45 K | 70 K | 300 K |
|---|---|---|---|---|---|
| Space Group | Cmcm | Cmcm | Cmcm | Cmcm | Cmcm |
| a (Å) | 11.1808(3) | 11.1808(3) | 11.1811(3) | 11.1808(3) | 11.1875(3) |
| b (Å) | 7.4990(2) | 7.4992(2) | 7.5003(2) | 7.5022(2) | 7.5235(2) |
| c (Å) | 7.6127(2) | 7.6125(2) | 7.6122(2) | 7.6122(2) | 7.6168(2) |
| V (Å)$^3$ | 638.29(4) | 638.28(4) | 638.37(4) | 638.52(4) | 641.10(4) |
| $R_{wp}$ | 4.23% | 4.20% | 4.16% | 4.17% | 3.81% |
| $\chi^2$ | 3.420 | 3.450 | 3.386 | 3.446 | 2.913 |
| Os-O1 (×2) (Å) | 1.98783(5) | 1.98765(5) | 1.98791(5) | 1.98784(5) | 1.99131(4) |
| Os-O2 (×4) (Å) | 1.96692(4) | 1.96648(4) | 1.96533(4) | 1.96501(4) | 1.96276(3) |
| ∠Os-O1-Os | 146.438(1) | 146.460(1) | 146.399(1) | 146.409(1) | 145.979(1) |
| La2 x | 0.2774(1) | 0.2773(1) | 0.2774(1) | 0.2773(1) | 0.2778(1) |
| La2 y | 0.3116(2) | 0.3119(2) | 0.3117(2) | 0.3117(2) | 0.3116(2) |
| O1 y | 0.4235(4) | 0.4235(4) | 0.4234(4) | 0.4234(4) | 0.4226(4) |
| O2 x | 0.1236(1) | 0.1236(1) | 0.1236(1) | 0.1237(1) | 0.1235(1) |
| O2 y | 0.1820(3) | 0.1819(2) | 0.1817(2) | 0.1815(2) | 0.1808(2) |
| O2 z | 0.0403(2) | 0.0405(2) | 0.0404(2) | 0.0404(2) | 0.0406(2) |
| O3 x | 0.3695(2) | 0.3694(2) | 0.3696(2) | 0.3694(2) | 0.3693(2) |
| O3 y | 0.0310(2) | 0.0311(3) | 0.0313(3) | 0.0309(3) | 0.0304(3) |
| La1 U | 0.0052(3) | 0.0050(3) | 0.0052(3) | 0.0061(3) | 0.0103(4) |
| La2 U | 0.0021(2) | 0.0021(2) | 0.0023(2) | 0.0026(2) | 0.0057(2) |
| Os U | 0.0019(2) | 0.0019(2) | 0.0023(2) | 0.0022(2) | 0.0045(2) |
| O1 $U_{eq}$ | 0.0043(6) | 0.0039(6) | 0.0044(6) | 0.0044(6) | 0.0077(7) |
| O2 $U_{eq}$ | 0.0092(4) | 0.0089(4) | 0.0095(3) | 0.0099(4) | 0.0154(4) |
| O3 $U_{eq}$ | 0.0043(4) | 0.0041(4) | 0.0044(4) | 0.0045(4) | 0.0072(5) |
| Os moment ($\mu_B$) | 1.2(2) | 1.3(2) | 0.8(2) | - | - |
| Moment angle | 30(8) | 31(9) | 55(13) | - | - |

TABLE 2 – Refinement parameters from neutron powder diffraction from $La_{2.8}Ca_{0.2}OsO_7$. $U_{eq}$ is defined as a third of the trace of the tensor for anisotropically refined U's, which are available as Supplemental Material. Moment angle is defined as the angle away from the *a* axis within the *ab* plane. Special positions used are La1 (0, 1/2, 0), La2 (x, y, 1/4), Os (0, 0, 0), O1 (1/2, y, 1/4), and O3 (x, y, 1/4).

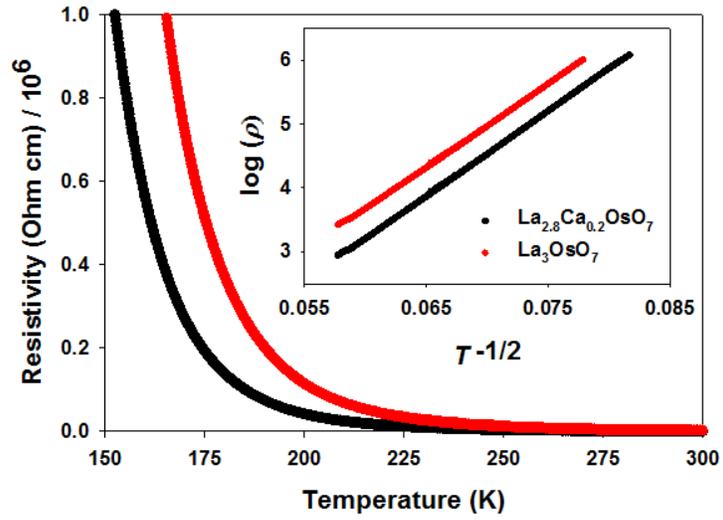

FIG. 3 (color online). The temperature dependence of the electrical resistivity of $La_3OsO_7$ (black) and $La_{2.8}Ca_{0.2}OsO_7$ (red). The inset shows the linear relationship on a $T^{-1/2}$ scale as described in the text.

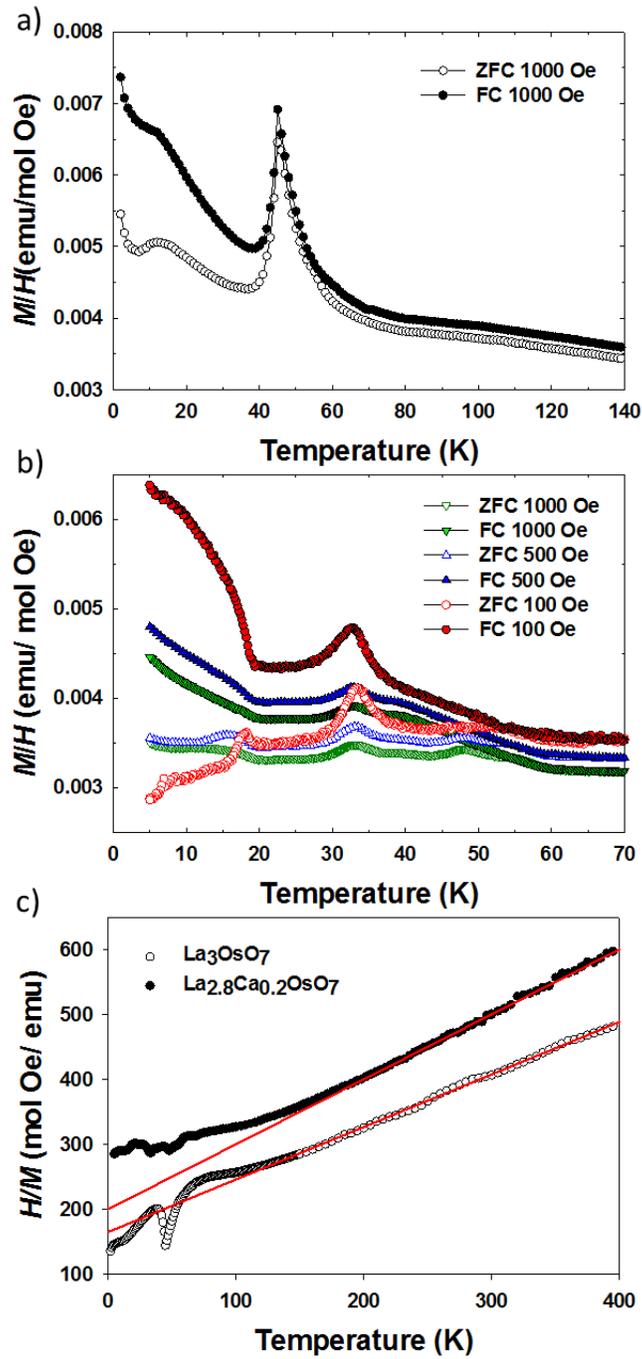

FIG. 4 (color online). The temperature dependence of the magnetization of (a) $La_3OsO_7$ and (b) $La_{2.8}Ca_{0.2}OsO_7$. The inverse relationship with Curie-Weiss fits (red lines) are shown in (c).

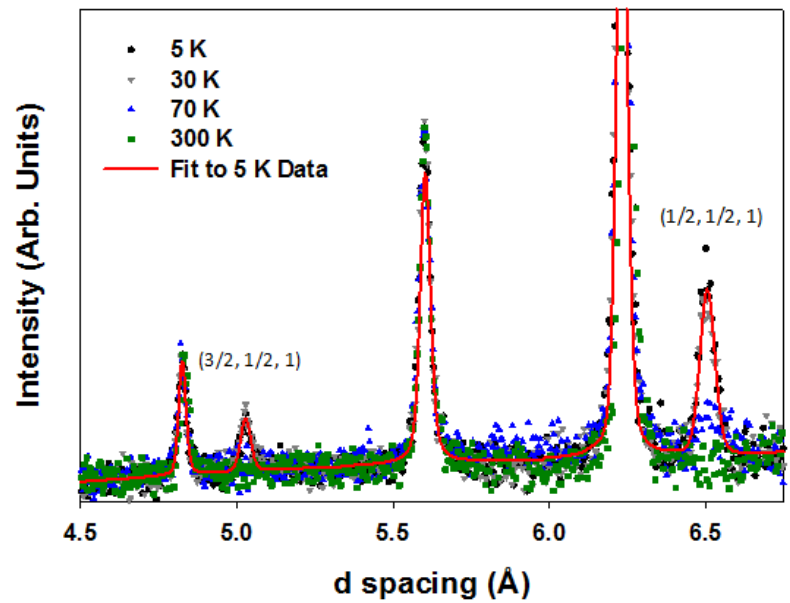

FIG. 5 (color online). High d-spacing neutron powder diffraction data showing the reflections arising due to magnetic scattering from $La_3OsO_7$. The calculated pattern including the magnetic model at 5 K is shown as the red curve.

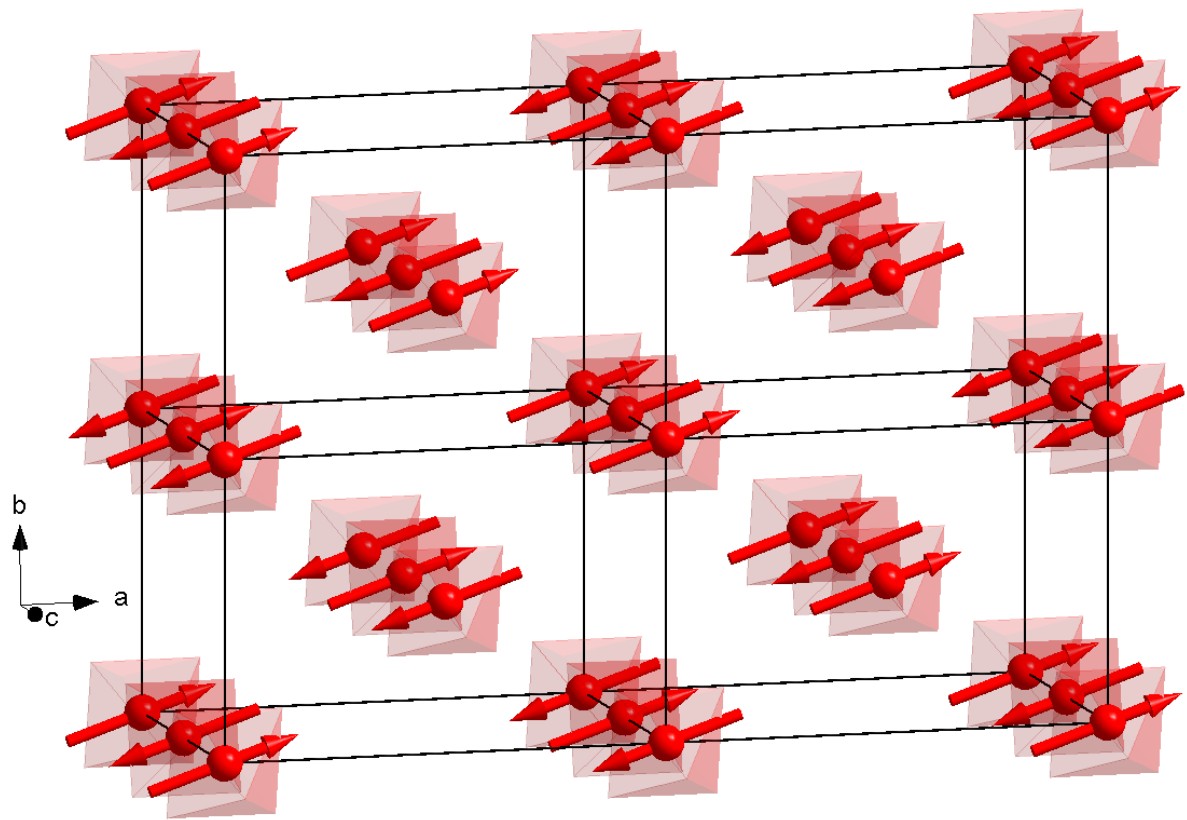

FIG. 6 (color online). Magnetic structure of $La_3OsO_7$ determined by analysis of neutron powder diffraction data shown on a 2 x 2 x 1 cell. The spins are coupled antiferromagnetically along each corner connected 1D chain of $Os^{5+}$ octahedra down the $c$ axis.